\documentclass[a4paper,fleqn,usenatbib]{mnras}

\usepackage[T1]{fontenc}
\usepackage{ae,aecompl}

\usepackage{graphicx}	
\usepackage{amsmath}	
\usepackage{amssymb}	

\title[A close encounter of the massive kind]{A close encounter of the massive kind\thanks{Based on observations obtained with the Hubble Space Telescope under
GO programs 10\,205, 10\,602, and 10\,898 and at the European Southern Observatory under programs 094.C-0397, 506.D-0495, and 297.D-5038.}}

\author[J. Ma{\'\i}z Apell\'aniz et al.]{
J. Ma{\'\i}z Apell\'aniz,$^{1}$\thanks{E-mail: jmaiz@cab.inta-csic.es}
H. Sana,$^{2}$
R. H. Barb\'a,$^{3}$
J.-B. Le Bouquin,$^{4}$
and R. C. Gamen$^{5}$
\\
$^{1}$Centro de Astrobiolog\'{\i}a. CSIC-INTA. Campus ESAC. Camino bajo del castillo s/n. E-28\,692 Villanueva de la Ca\~nada. Madrid. Spain.\\
$^{2}$Institute of Astronomy. KU Leuven. Celestijnenlaan 200 D. B-3001 Leuven. Belgium.\\
$^{3}$Departamento de F\'{\i}sica y Astronom\'{\i}a. Universidad de La Serena. Av. Cisternas 1200 Norte. La Serena. Chile.\\
$^{4}$Institut de Plan\'etologie et d'Astrophysique de Grenoble (IPAG) UMR 5274. F-38\,041. Grenoble. France.\\
$^{5}$Instituto de Astrof\'{\i}sica de La Plata (CONICET, UNLP). Paseo del Bosque s/n. 1900 La Plata. Argentina.
}

\date{Accepted 2016 October 7. Received 2016 September 7; in original form 2016 August 16}

\pubyear{2016}

\begin{document}
\label{firstpage}
\pagerange{\pageref{firstpage}--\pageref{lastpage}}
\maketitle

\begin{abstract}
We have used (a) HST ACS imaging and STIS spectroscopy, (b) ground-based PIONIER/VLT long-baseline interferometry, and (c) ground-based spectroscopy from different 
instruments to study the orbit of the extreme multiple system HD~93\,129~Aa,Ab, which is composed of (at least) two very massive stars in a long-period orbit with $e>0.92$ 
that will pass through periastron in 2017/2018. In several ways, the system is an $\eta$~Car precursor. Around the time of periastron passage the two very strong winds will 
collide and generate an outburst of non-thermal hard X-ray emission without precedent in an O+O binary since astronomers have been able to observe above Earth's atmosphere. 
A coordinated multiwavelength monitoring in the next two years will enable a breakthrough understanding of the wind interactions in such extreme close encounters. 
Furthermore, we have found evidence that HD~93\,129~Aa may be a binary system itself. In that case, we could witness a three-body interaction that may yield a runaway star 
or a stellar collision close to or shortly after the periastron passage. Either of those outcomes would be unprecedented, as they are predicted to be low-frequency events in 
the Milky Way. 
\end{abstract}

\begin{keywords}
ephemerides -- binaries: visual -- stars: individual: HD~93\,129~Aa,Ab -- 
stars: kinematics and dynamics -- stars: winds, outflows -- X-rays: stars
\end{keywords}



\section{Introduction}

$\,\!$\indent Massive stars love company, with the vast majority of them (if not all) being born in multiple systems of two or more 
stars \citep{Masoetal98,Sanaetal12a,Sotaetal14,Sanaetal14}. Many of those systems are in low-eccentricity, short-period (1 day to a few months) orbits that 
lead to slow, prolonged interactions that play an important role in their evolution \citep{Pols94}. Recent studies have discovered a population of eccentric 
systems with periods around 1 year \citep{SimDetal15a,Maizetal15a}, of which the most extraordinary case is $\eta$~Car, a high-eccentricity ($e=0.9$)
binary in a 5.54-year orbit \citep{Dami96,Damietal00} where the periastron passages have led to violent events that have changed the fate of the 
system in short time scales \citep{Smit11a}.

HD~93\,129~A, the central system of the compact massive young cluster Trumpler~14 in the Carina Nebula, has long been recognized as one of the hottest and
most massive stellar systems in the Galaxy \citep{Walb71b}. It is the prototype of the earliest O spectral subtype of supergiants (O2~If*) and its only 
example known in the Galaxy \citep{Walbetal02b,Sotaetal14}. \citet{Nelaetal04} resolved it with the Fine Guidance Sensor (FGS) on board the
Hubble Space Telescope (HST) as an astrometric binary (Aa,Ab components). \citet{Maizetal08b} detected a relative astrometric motion 
between the two components, with the separation in the plane of the sky between them decreasing in a nearly inward direction, and suggested that the system
followed a highly eccentric and/or inclined orbit. The speed of the inward motion increased in subsequent years, lending credence to the high eccentricity
option \citep{Sanaetal14,Benaetal15}. The previously published data show a decrease from a distance of 66 milliarcseconds (mas) in 1996 to 28 mas in 2012
(2.4 mas/a on average).

\section{Methods}


\subsection{Astrometric data}

$\,\!$\indent We measured the relative astrometric orbit of HD~93\,129~Aa,Ab using positions from:

\begin{itemize}
 \item One previously published epoch from FGS/HST \citep{Maizetal08b,Benaetal15}.
 \item Two epochs from the High Resolution Camera (HRC) of the Advanced Camera for Surveys (ACS) onboard HST \citep{Maizetal08b}.
 \item Five epochs from the PIONIER four-beam combiner at the Very Large Telescope and Interferometer (VLTI) at the European Southern Observatory (ESO) 
       \citep{LeBoetal11,Sanaetal14}. Of those, only the first one had been published before.
\end{itemize}

The measured values are listed in Table~\ref{relastrom}.

\begin{table*}
\caption{Relative astrometry for the HD~93\,129~Aa,Ab system. The first three columns show the Julian Date of the observation, the separation,
and the position angle (measured counterclockwise from North). The next three columns give the semimajor and semiminor axes plus the position angle of 
the uncertainty ellipses. The last column gives the instrument used.}
\centerline{
\begin{tabular}{ccrrrrc}
JD             & separation  & \multicolumn{1}{c}{PA} & $\sigma_{\rm max}$ & $\sigma_{\rm min}$ & PA$_\sigma$ & instrument   \\
(d)            & (mas)       & (deg)                  & (mas)              & (mas)              & (deg)       &              \\
\hline
2\,450\,214.42 &  66.00      &   11.50                & 13.82              & 2.00               & 102         & FGS/HST      \\
2\,453\,224.09 &  53.02      &   14.83                &  1.30              & 1.16               & 105         & HRC/HST      \\
2\,453\,950.06 &  45.11      &   16.75                &  1.42              & 1.04               &  17         & HRC/HST      \\
2\,456\,087.98 &  28.35      &    9.57                &  1.42              & 0.73               & 149         & PIONIER/VLTI \\
2\,456\,995.36 &  17.49      &    5.98                &  1.50              & 0.59               & 132         & PIONIER/VLTI \\
2\,457\,392.23 &  11.61      &    2.66                &  0.65              & 0.29               & 130         & PIONIER/VLTI \\
2\,457\,481.24 &  10.22      &    0.87                &  0.95              & 0.24               &  51         & PIONIER/VLTI \\
\hline
\end{tabular}
}
\label{relastrom}
\end{table*}

\begin{figure*}
\centerline{\includegraphics[width=0.32\linewidth]{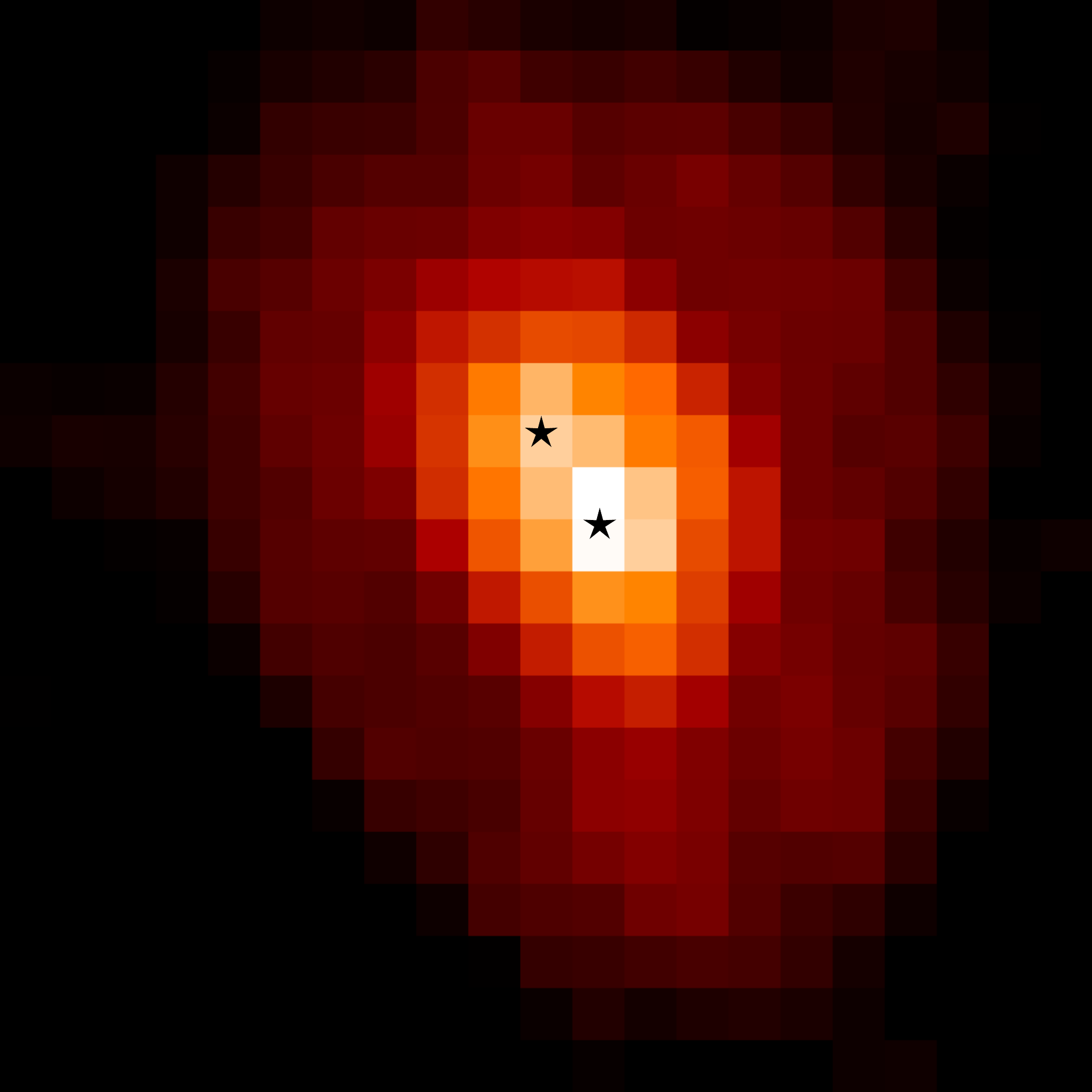} \
            \includegraphics[width=0.32\linewidth]{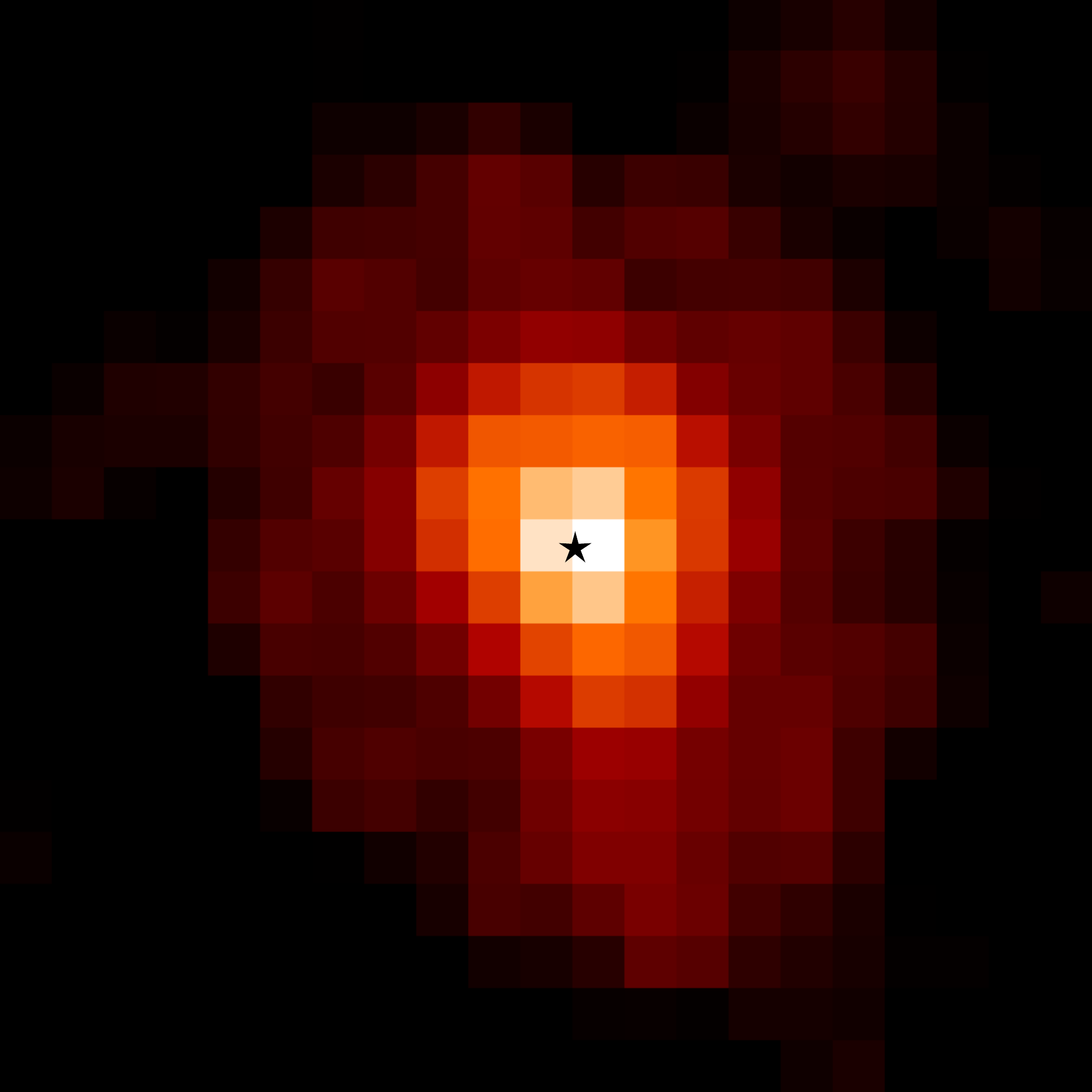} \
            \includegraphics[width=0.32\linewidth]{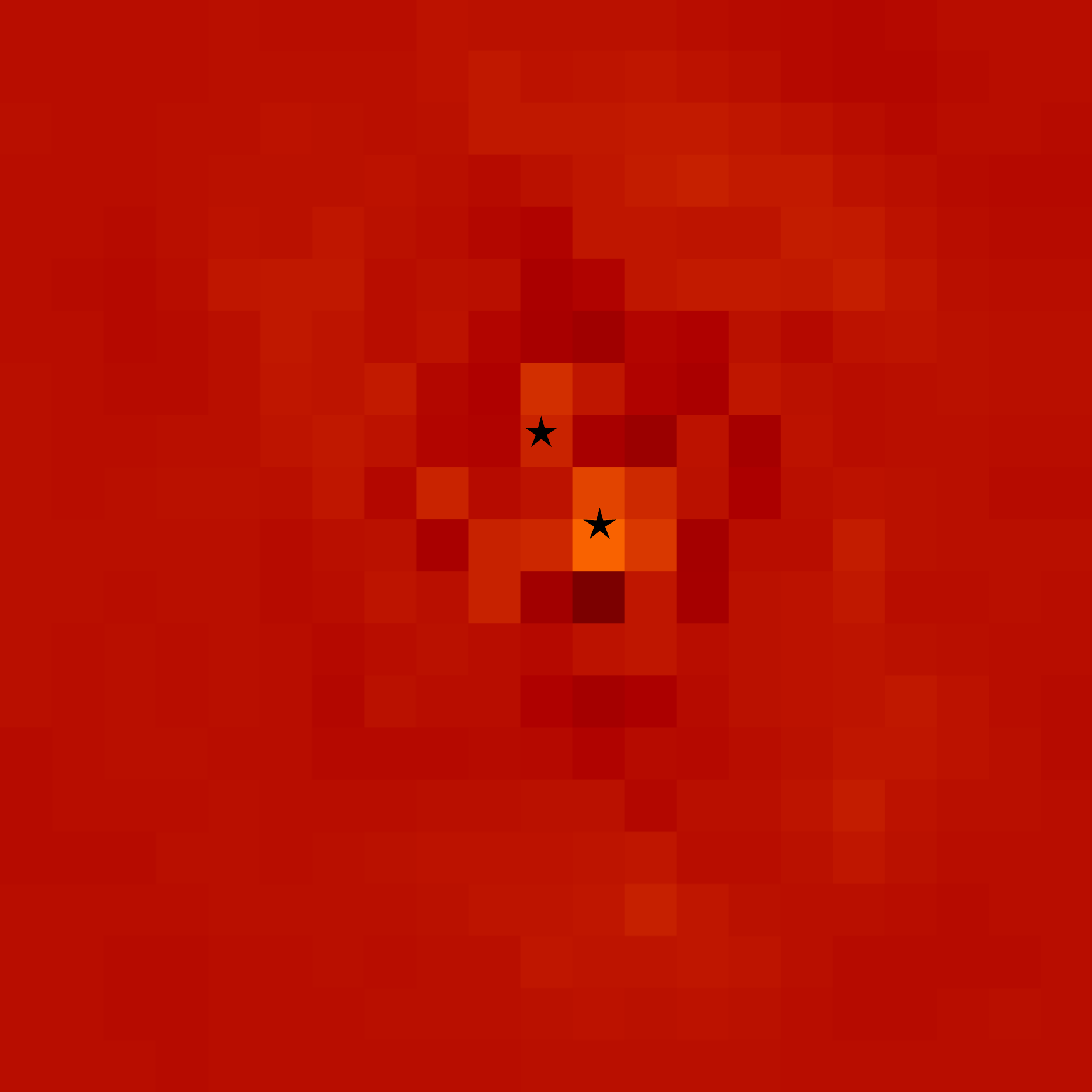}}
\caption{[left] Imaging data, [center] Point-Spread Function (PSF), and [right] fit residual (data with the two stars subtracted) for one of the F435W HRC 2004
exposures. In the left and center panels the intensity scale is logarithmic between 0.1\% and 100\% of the maximum while in the right panel the intensity
scale is linear between -3\% and 3\% of the data maximum. The star symbols mark the position of the two components (Aa, center, and Ab, upper left) in the 
two extreme panels and the PSF center in the center one. The field size is 0\farcs53$\times$0\farcs53.}
\label{HRC}
\end{figure*}

The HRC data (from HST GO programs 10\,205, 10\,602, and 10\,898) were processed using a 
crowded-field photometry package created by the lead author that works on the geometrically distorted (not on the drizzled) exposures \citep{AndeKing04}. Three 
different HRC filters (F330W, F435W, and F850LP) were used. The excellent quality of the double PSF fit can be appreciated in Fig.~\ref{HRC}. The existence
of seven unsaturated bright stars belonging to Trumpler~14 in the HRC field allowed us to measure not only relative motions (in the plane of the sky) but also 
absolute motions by assuming that the rest of the stars remained stationary during the two-year span between the two HRC epochs (2004.589 and 2006.586). The F435W
and F850LP filters were used for the absolute proper motion measurement.

The new VLTI/PIONIER observations were obtained with the auxiliary telescopes in configuration A0-G1-J2-J3, which provides a maximum baseline of 132~m, 
corresponding to a spatial resolution element in the H-band of 2.5~mas. The observations were processed following the procedure described 
in \citet{Sanaetal14} and references therein. The HD~93\,129~Aa,Ab binary was clearly resolved at each epoch. A binary model with three 
parameters (separation, position angle and brightness ratio) was fitted to the visibilities and closure phases in order to obtain sub-milliarcsec precision relative 
astrometry of the two components of the system.

\subsection{Spectroscopic data}

\begin{figure*}
\centerline{\includegraphics[height=8.5cm]{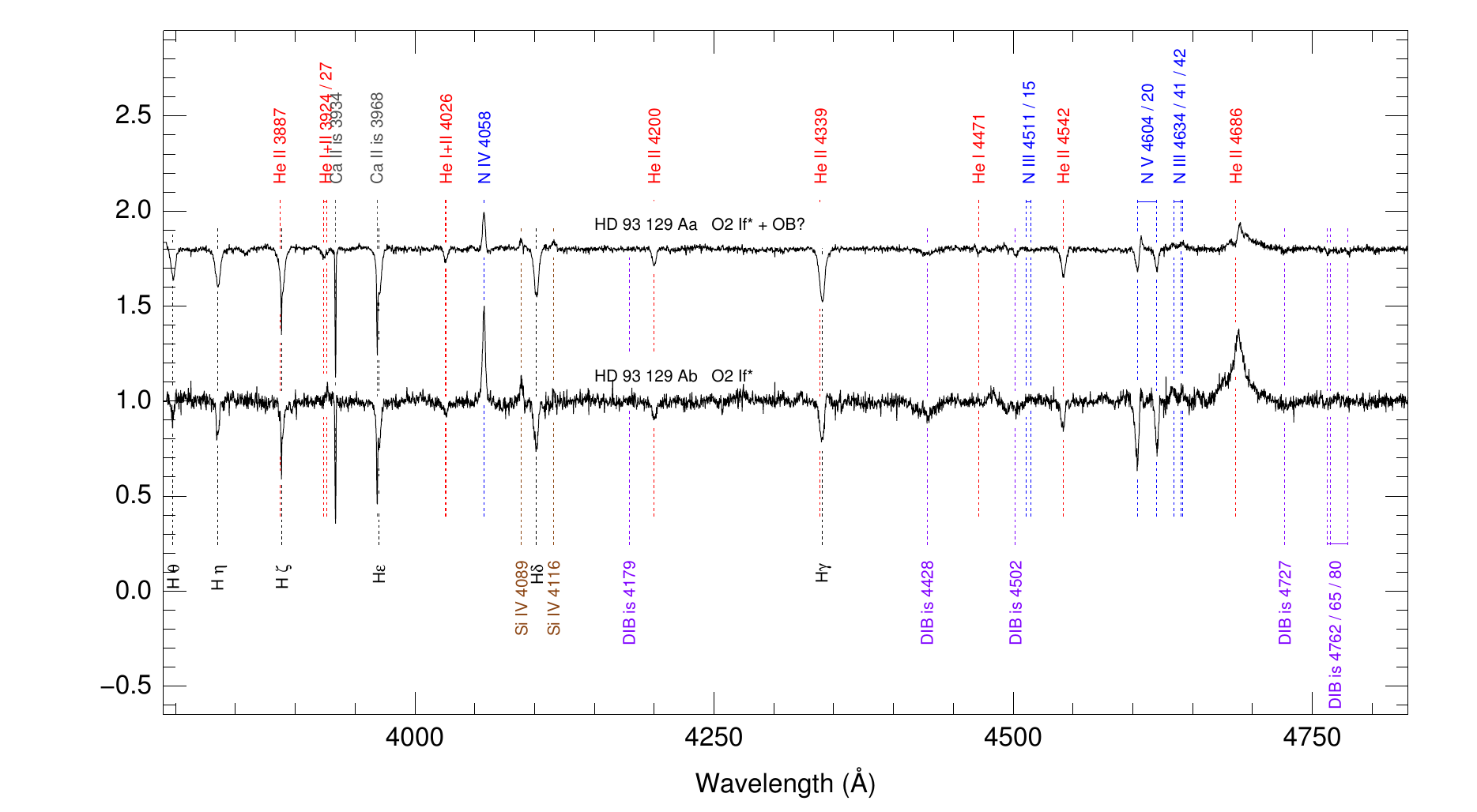} \
            \includegraphics[height=8.5cm]{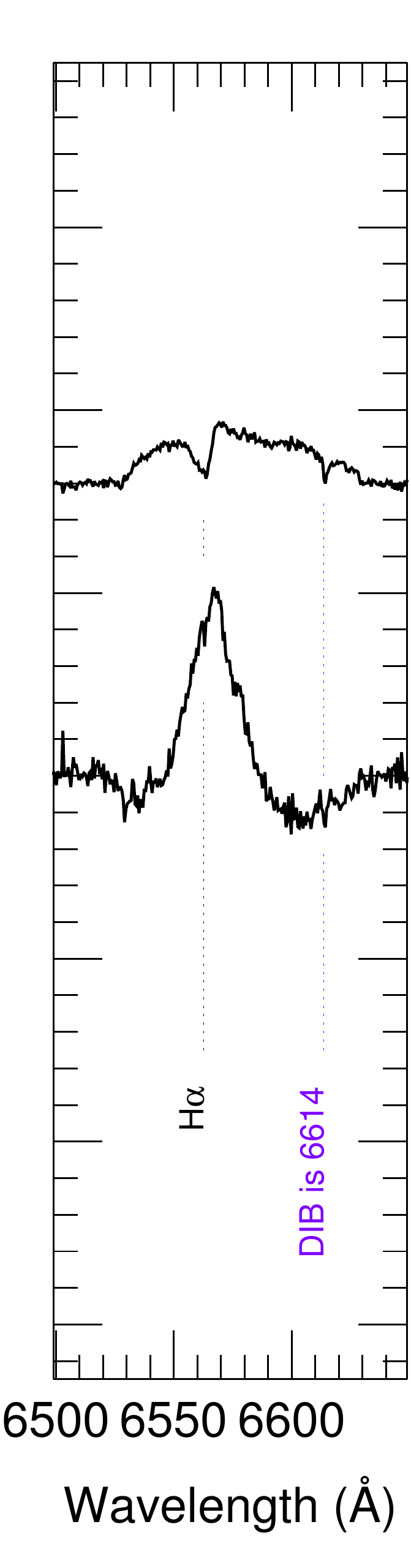}}
\caption{Spatially-resolved rectified STIS spectra of HD~93\,129~Aa and Ab: [left] blue-violet region and [right] H$\alpha$. The spectra have been displaced in the vertical
direction for clarity. The HD~93\,129~Ab spectrum is noisier than that of Aa due to the $\approx$1~magnitude difference between the two components (see also 
Fig.~\ref{HRC}, note that the intensity scale in the left and center panels there is logarithmic).}
\label{spectra}
\end{figure*}

$\,\!$\indent We observed HD~93\,129~Aa,Ab with the Space Telescope Imaging Spectrograph (STIS) onboard HST using the G430M and G750M gratings in 2010.264 aligning the 
slit with the position angle of the system at that time, when the pair was separated by 36 mas, to obtain individually resolved spectra of the two components 
(Fig.~\ref{spectra}). The separation between the stars was smaller than the spatial pixel size of the STIS CCD (50 mas). To overcome that, we [a] dithered the 
exposures spatially with a 2-point pattern in order to decrease the effective pixel size to 25 mas and [b] we used the techiques developed for MULTISPEC \citep{Maiz05a} 
to obtain spatially-resolved spectra for both the Aa and Ab components.

We have obtained 42 optical spectra of HD~93\,129~Aa,Ab between 2005 and 2016 using the FEROS spectrograph attached at the MPE/ESO-2.2 m telescope (La Silla 
Observatory, Chile) and the {\'E}chelle spectrograph attached to the du Pont 2.5 m (Las Campanas Observatory, Chile). The data were obtained within the framework of 
the OWN Survey \citep{Barbetal10}. The slit widths used in those observations (2\arcsec\ in FEROS, and 1\arcsec\ in {\'E}chelle) do not allow us to spatially resolve the 
Aa and Ab components. They provide nominal resolving powers of 48\,000 and 45\,000, respectively. The data were processed using the FEROS Data Reduction System implemented
within ESO-MIDAS software, and the {\'e}chelle package in IRAF. Additional processing such as rectification and telluric corrections were performed using IRAF. The final 
signal-to-noise ratio of the one-dimensional extracted spectra depends on the wavelength and the observing night, varying between 100 and 400. We have measured the 
radial velocity, equivalent width, and full width at half maximum for a number of absorption and emission lines in the spectra in order to determine the behavior of 
those parameters with time. 

UVES/VLT observations were obtained on the night of June 27, 2016 with the DIC2 436+760 and 346 + 820 setups, yielding a continuous wavelength coverage from 305~nm to 
498~nm and from 570~nm to 945~nm. Two back-to-back exposures of 150~s were obtained in each setup. An entrance slit of 0\farcs6 was used for both arms, 
yielding a nominal spectral resolving power of 60\,000. The data were processed in the standard way using the ESO CPL pipeline v5.7.0 under the REFLEX environment 
and calibration frames obtained within 12~h of the scientific observations. The extracted 1-dimensional spectra were rectified by fitting a low order polymial 
through the continuum and show a signal-to-noise ratio per resolution element of about 400. To complement our time series, archival UVES observations of 
HD~93\,129~Aa,Ab from April 2013 and April 2015 were further retrieved from the ESO archive data base and processed in a similar way.

\section{Results}

\subsection{Relative astrometric orbit}

\begin{figure}
\centerline{\includegraphics[width=0.95\linewidth]{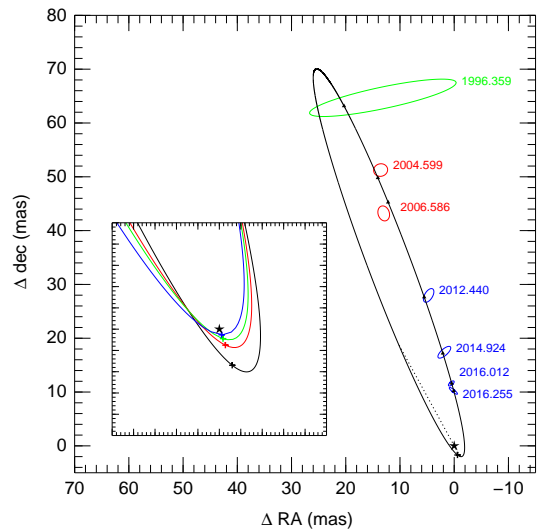}}
\caption{Best (mode) relative two-body orbit projected on the plane of the sky for HD~93\,129~Aa,Ab based on the astrometric data. The star shows the orbital 
center, the ellipses the data (green=FGS, red=HRC, blue=PIONIER), the text the observation epochs (in a), the triangles the predicted positions, the 
cross the periastron location, and the dotted line the line of nodes. The inset shows the inner 10~mas$\times$10~mas with four different orbits with 
$\chi^2_{\rm red} < 1$. The mode orbit is drawn in black and is the one with the latest periastron epoch (2018.14) and distance (2.085 mas) and the lowest 
eccentricity (0.952). The other three orbits are drawn in red, green, and blue and follow decreasing sequences in periastron epoch (2017.66, 2017.48, and 
2017.30) and distance (0.975, 0.615, and 0.375 mas) and increasing sequences in eccentricity (0.980, 0.988, and 0.993). Crosses mark the periastron locations. 
1~mas~=~2.35~AU at the distance to the Carina Nebula \citep{Smit06a}.}
\label{orbit}
\end{figure}

\begin{figure*}
\centerline{\includegraphics[width=0.52\linewidth, bb=28 28 566 566]{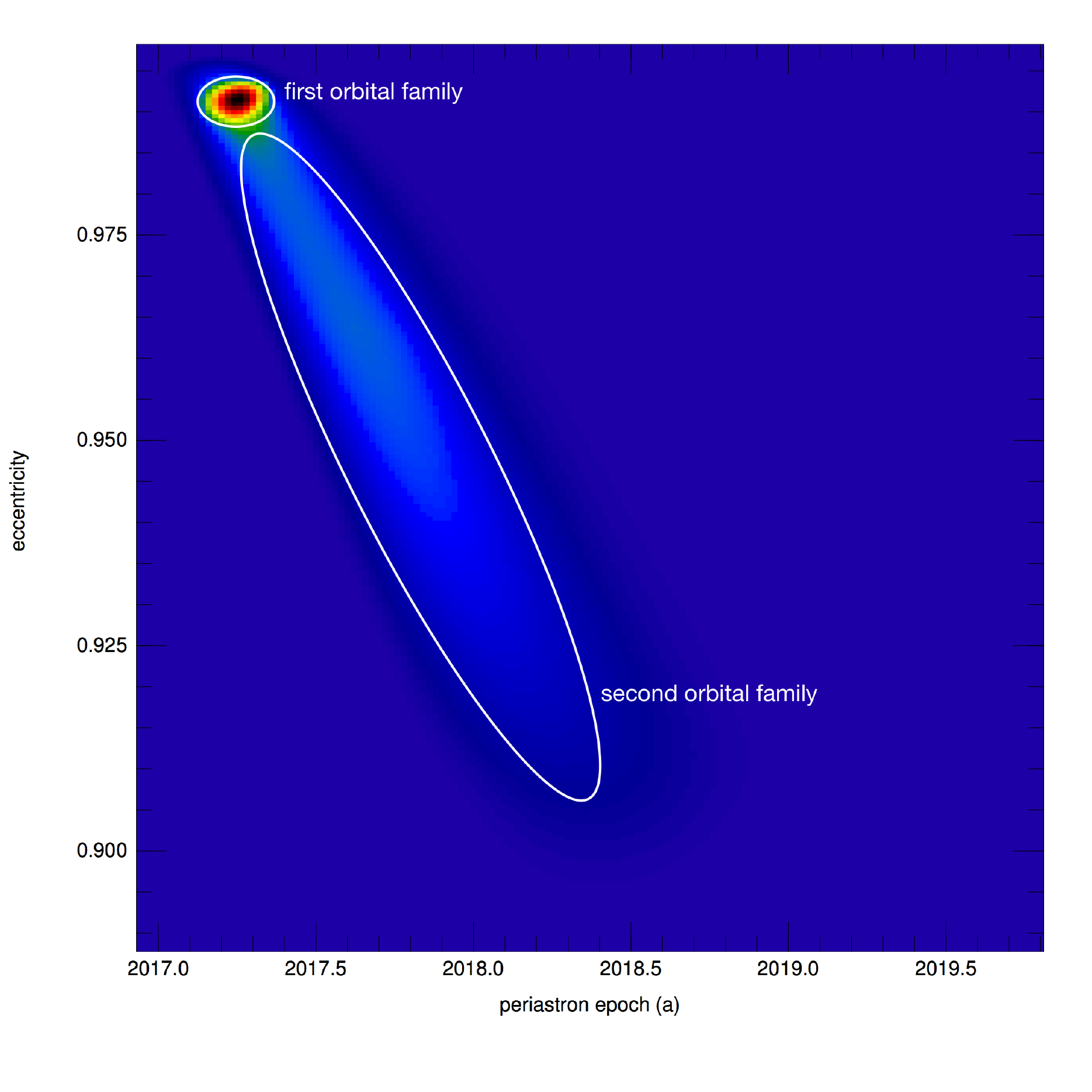} \
            \includegraphics[width=0.52\linewidth, bb=28 28 566 566]{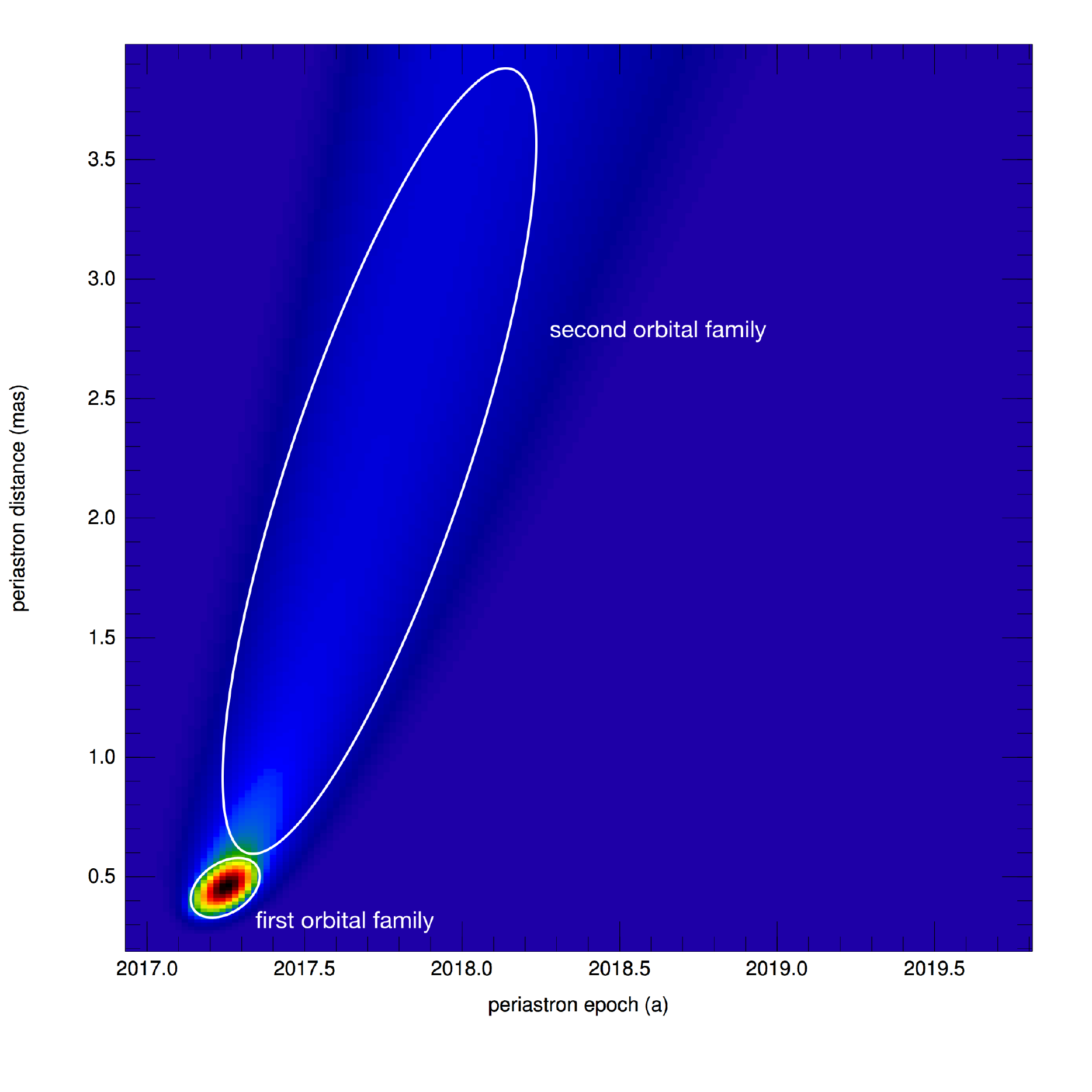}}
\caption{Likelihood distribution for the relative astrometric two-body orbit collapsed into the $T_0$-$e$ [left] and $T_0$-$d$ [right] planes. The regions with a 
large likelihood have been grouped into two orbital families (see text) and ellipses have been drawn to guide the eye. Note that each of the two families has relatively
complex shapes in seven-dimensional space (i.e. they cannot be simply described by seven-dimensional ellipsoids). The integrated likelihood of the two families is similar,
as the highest peak values (as projected in these planes) of the first one are compensated by the larger volume occupied by the second one.}
\label{likelihood}
\end{figure*}

\begin{figure}
\centerline{\includegraphics[width=\linewidth]{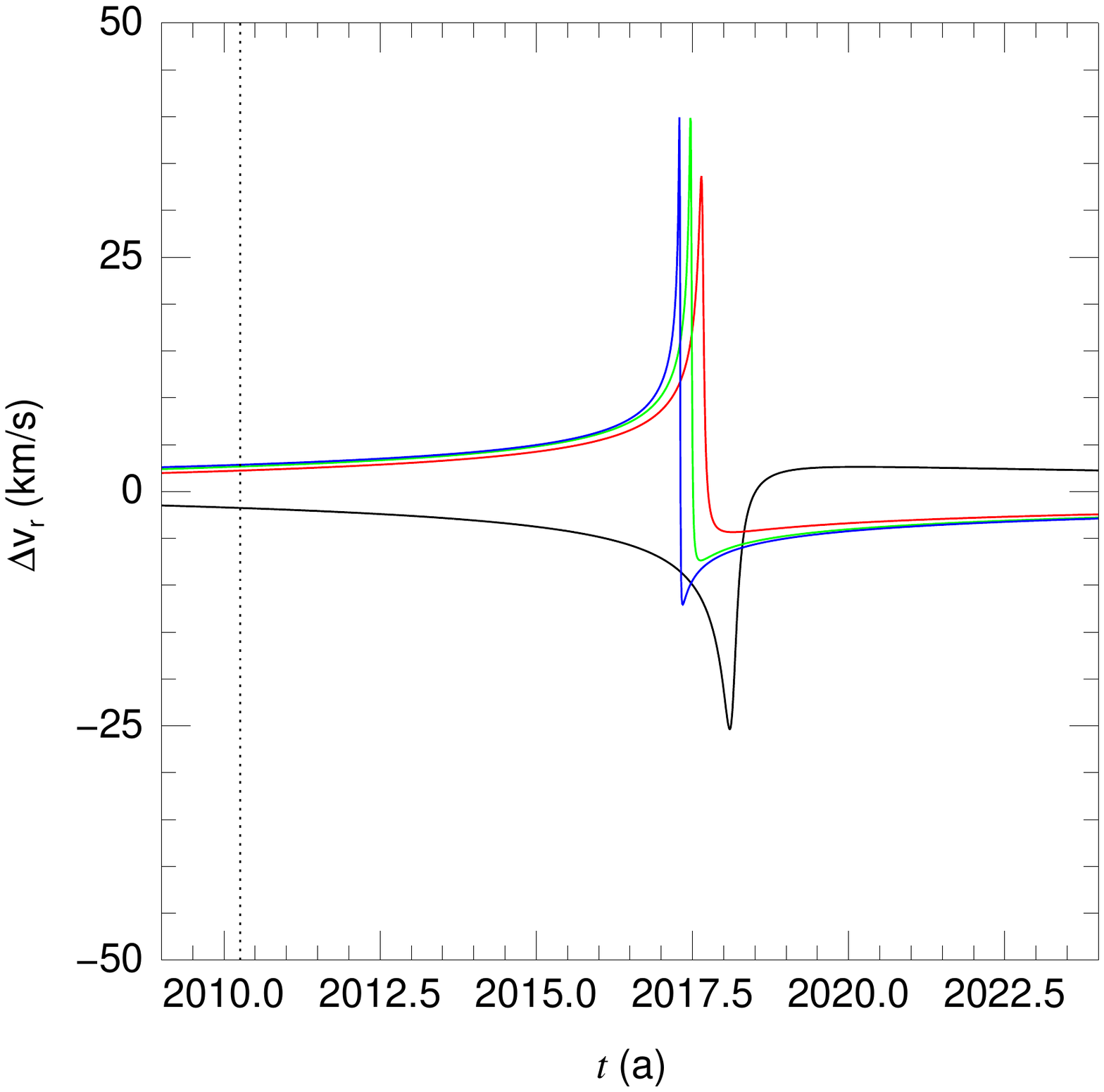}}
\caption{Radial velocity predictions for the four relative astrometric two-body orbits shown in the inset of Fig.~\ref{orbit}. The dotted vertical line marks the 
epoch of the STIS observation. Note that currently $\Omega$ is undefined by $\pm$180 degrees, which causes the existence of solutions where Ab will be moving away 
from us near periastron and solutions where Ab will be moving towards us at that time. We show three of the former and one of the latter.}
\label{vrad}
\end{figure}

$\,\!$\indent Since only a fraction of the relative astrometric orbit is available and only seven epochs have been obtained so far (Table~\ref{relastrom}), 
it was not possible to derive precise values for all seven independent parameters that need 
to be determined, which we chose to be (among the different possible combinations): [1] orbital period $P$, [2] periastron epoch $T_0$, [3] eccentricity $e$, 
[4] periastron distance $d$, [5] inclination $i$, [6] ascending node longitude $\Omega$, and [7] the difference between the periastron argument and the ascending 
node longitude $\omega-\Omega$. A simple search for the maximum likelihood point (mode) under such circumstances is not sufficient, as the likelihood 
in the 7-dimensional parameter space is likely to have multiple peaks. For that reason, we conducted a full search over the plausible parameter ranges, we 
calculated the likelihood at $\sim 10^{11}$ grid points, and we selected the $2.6\times 10^8$ points with the highest values. Results are shown in Table~\ref{parameters}
and in Figs.~\ref{orbit},~\ref{likelihood},~and~\ref{vrad}.

\begin{table}
\caption{Relative two-body orbital parameters for HD~93\,129~Aa,Ab based on the astrometric data. The first block lists the fitted parameters and the second block
the derived ones. The very large uncertainties for $\Omega$ and $\omega$ arise from the existence of two subfamilies of solutions separated by 180 degrees (i.e. we do not 
know if the Ab component is moving towards us or away from us at a given point).}
\centerline{
\begin{tabular}{lcc}
quantity        & units     & value                     \\
\hline
$P$             & a         & $121^{+60}_{-39}$         \\[1ex]
$T_0$           & a         & $2017.60^{+0.38}_{-0.32}$ \\[1ex]
$e$             &           & $0.967^{+0.023}_{-0.026}$ \\[1ex]
$d$             & mas       & $2.7^{+2.2}_{-1.7}$       \\[1ex] 
$i$             & deg       & $117^{+28}_{-7}$          \\[1ex]
$\Omega$        & deg       & $192^{+15}_{-173}$        \\[1ex]
$\omega-\Omega$ & deg       & $-24^{+12}_{-12}$         \\[1ex]
\hline
$a_{\rm Aa,Ab}$ & mas       & $50^{+12}_{-7}$           \\[1ex]
$\omega$        & deg       & $177^{+153}_{-143}$       \\[1ex]
$m_{\rm Aa,Ab}$ & M$_\odot$ & $108^{+52}_{-15}$         \\[1ex]
\hline
\end{tabular}
}
\label{parameters}
\end{table}

Two families of orbital solutions are observed\footnote{In turn, each family is divided into two subfamilies separated by 180~degrees in $\Omega$. This degeneracy is caused 
by the absence of radial velocity points in the astrometric orbit that would allow us to decide whether the secondary is moving towards us or away from us at a given point in
the orbit.}. The first one has clearly defined values of $T_0 \approx 2017.25$~a, 
$e\approx 0.991$, and $d \approx 0.45$~mas and more loosely constrained values of $P$ between 100~a and 250~a and of $i$ between 150~degrees and 180~degrees. The 
second one is more dispersed in the parameter space, with $P$ between 50~a and 120~a, $T_0$ between mid-2017 and mid-2018, $e$ between 0.92 and 0.99, $d$ between 
0.5~mas and 4~mas, and $i$ between 100~degrees and 130~degrees. In any case, all solutions are highly eccentric and with periastron passages in 2017/2018 within 
less than 14 AU (6~mas). Further VLTI observations are needed to establish which of the two families is the right one. No periastron passage between two massive stars 
with such a high eccentricity has ever been observed.

\subsection{Spectral classification}

$\,\!$\indent The spatially resolved STIS spectroscopy allowed us to obtain individual spectral classifications for the Aa and Ab components.
Both stars have similar spectral types, with very weak or no He\,{\sc i}~4471 absorption, strong He\,{\sc ii}~4542 and N\,{\sc v}~4604+20 
absorptions, He\,{\sc ii}~4686 in emission, and strong N\,{\sc iv}~4058 emission (much larger that the weak N\,{\sc iii}~4634+41+42 emission). This makes both
objects O2~If*, the earliest subtype of O supergiants \citep{Walbetal02b,Sotaetal11a}, and contrasts with previous assumptions that the secondary was of a later spectral 
type such as O3.5 (note that such claims were based on indirect evidence, not on observed spectra). Even so, there are some differences between the two components: 
[a] H$\alpha$ and He\,{\sc ii}~4686 are in pure emission for Ab but have signs of an absorption component for Aa and [b] He\,{\sc i}~4471 absorption is completely absent 
for Ab but is seen (very weak) in Aa (the weak absorption feature is also observed in the spatially unresolved ground-based spectra). Those differences could be explained 
if Aa itself is a tight binary (Aa1,Aa2) composed of an O2 supergiant (Aa1) and a later-type companion (Aa2). See below for further evidence regarding this hypothesis.

\subsection{Mass ratio}

$\,\!$\indent The absolute proper motion derived from the HRC/ACS data yield a
mass ratio between the Ab and Aa components of 0.49$\pm$0.22. This is anomalously small for two objects of the same spectral type but could be at least partially 
explained if Aa is indeed a tight binary, as suggested above. The relative astrometric orbit also yields an estimate for the total mass for the system
(Table~\ref{parameters}). While uncertainties are still large, given that we only have a partial orbit and that two families of solutions are allowed, we obtain values 
firmly above 60~M$_\odot$ for Aa and above 30~M$_\odot$ for Ab (and possibly much higher), making HD~93\,129~Aa,Ab the most massive astrometric binary (or triple 
system) known \citep{Sanaetal13a,SanBetal13}.

\subsection{The spectroscopic multiplicity of the HD~93\,129~A system}

$\,\!$\indent The spatially resolved STIS spectra also allowed us to measure the line-of-sight velocity difference $\Delta v_r$ between Ab and Aa (positive for Ab 
moving away from us) in 2010.264. The N\,{\sc iv}~4058 emission line yielded $-$0.3$\pm$2.8~km/s (note that this line depends on the wind properties) and the 
N\,{\sc v}~4620 absorption line yielded $+$6.0$\pm$5.3~km/s. Both results are compatible and indicate that in 2010.264 there was little motion along the line of 
sight, which is consistent with the expected velocities from the astrometric results (Fig.~\ref{vrad}) but provides no strong additional constraint on the 
relative orbit (e.g. in breaking the degeneracy between the two $\Omega$ subfamilies)..

In the ground-based spectrocopic data (FEROS, {\'E}chelle, and UVES) Aa and Ab are unresolved (HD~93\,129~B, a more distant bright star is not included). We observe 
time-dependent velocity differences up to 15~km/s in the strong N\,{\sc v}~4620 absorption line and up to 50~km/s in the weak He\,{\sc i}~5876 absorption line (the former 
would originate in Aa1, the primary of the Aa1,Aa2 system, and in Ab while the latter would originate in Aa2, the secondary of the Aa1,Aa2 system), as well as profile 
variations in the He\,{\sc ii}~4686 emission line. Those changes occur in time scales of a few days and they are not secular, as would be expected from the evolution of the Aa,Ab 
orbit (Fig.~\ref{vrad}), so they represent further evidence that Aa is a tight binary. We are analyzing the modulation of these radial velocity variations. Some possible 
periods have been detected but a definitive orbit cannot be derived at this time. We plan to obtain further epochs to achieve that and we will report on the results as
soon as we obtain them, given the proximity of the Aa,Ab periastron passage.

\section{Analysis}


\subsection{HD~93\,129~A as a binary system}

$\,\!$\indent Former radio observations have revealed that the stellar winds of components Aa and Ab collide in-between the two stars. In this so-called wind-collision 
region (WCR), electrons are accelerated at the shock interfaces up to relativistic energies. As a result, the WCR produces intense non-thermal (synchrotron) radio 
emission \citep{DeBe07,DeBeRauc13}. The last radio observations, obtained in 2009 when the distance between the two stars was about two orders of magnitude larger 
than the value at periastron, showed that the flux was increasing with time (\citealt{Benaetal15}, whose wind parameters we use here). However, given the extremely close 
encounter that we predict, the WCR will be pushed deep into the radio sphere of the stars and the non-thermal radio emission from the WCR will now be entirely absorbed by 
the stellar winds, except maybe at the higher frequencies ($>$20~GHz). 

While the periastron passage of HD~93\,129~A will be radio-quiet, the situation will be very different in the X-ray (and possibly $\gamma$-ray) domain. If the distance at
periastron is indeed $\approx$0.45~mas, as predicted by our first family of orbital solutions,
the non-thermal hard X-ray emission produced by the wind-collision (mostly through inverse Compton scattering) should 
increase by two order of magnitudes compared to the epoch of the last X-ray measurements \citep{Gagnetal11}. This is about 10 times more intense that the most 
optimistic scenario so far \citep{DelPetal16}. Applying a 1/$D$ scaling to the existing models \citep{DelPetal16} to account for the new periastron distance (which we 
can do because the WCR cooling properties remains in the adiabatic regime, even at periastron), we find that the non-thermal emission will dominate the X-ray flux 
of the system above a couple of keV, possibly leading to one of the more obvious high-energy detections of non-thermal processes in stars. 
A characterization of the evolution of WCR non-thermal radio + X-ray properties and their relative comparison will allow astronomers to obtain unprecedented 
constraints on the particle acceleration mechanism, with immediate implications to understand the production mechanism at the origin of cosmic rays, which are 
thought to be accelerated through similarly strong shocks in supernovae remnants.

How does HD~93\,129~Aa,Ab compare to $\eta$~Car? The masses are not too different, especially if the low-mass models for $\eta$~Car are considered
(\citealt{Clemetal15,KashSoke16} and references therein), but the primary star in $\eta$~Car is more evolved and the 
interaction with the secondary appears to have become significant for their evolution just after the primary became an evolved hypergiant, thus allowing the 
companion to graze the hydrostatic core radius of the primary. That is still far in the future for HD~93\,129~Aa,Ab in terms of number of orbits (thousands or tens of 
thousands) so, in the absence of perturbations (see below), we should not expect mass ejections during the next periastron like those experienced by $\eta$~Car. 
One important difference exists between $\eta$~Car and HD~93\,129~Aa,Ab. $\eta$~Car is located in Trumpler~16, a 
loose cluster whose full extent is not clear, and has no rival there in terms of stellar mass. On the other hand, HD~93\,129~Aa,Ab is at the center of 
Trumpler~14, a compact, dense cluster with two other very massive stars in the core itself, HD~93\,129~B and HD~93\,128 \citep{Sotaetal14}. Therefore, the Aa,Ab highly 
eccentric orbit should be easy to perturb and the system may experience a very different demise (becoming unbound or having a too-close approach at periastron). In any 
case, a close monitoring of the orbit in the incoming years should shed some light on the issue.

Most known examples of very massive stars are of spectral type WNLh/ha or O Iafpe (also known as ``late Of/WN'' or ``cool slash'', \citealt{Sotaetal14}), see
e.g. \citet{Moffetal04,Crowetal10,CrowWalb11} and \citet{Maizetal15a} for a recent listing. However, given the small sample known it is not clear if all of the stars with
masses $\gtrsim 100$~M$_\odot$ have the WR features that prompted Phil Massey to call them ``Of stars on steroids'' \citep{MoffPuls03}. Our determination of 
the masses of the HD~93\,129~A stars currently has relatively large uncertainties but our current results do not exclude values around 100~M$_\odot$ for the most massive 
component. A better determination of the masses could make this system a crucial case to decide the issue, as HD~93\,129~A has the advantage over other examples of being
an astrometric binary, thus exempt from the inclination problem that affects the determination of masses in non-eclipsing spectroscopic binaries. But why should two stars of
the same mass have different spectral types? A traditional stellar-evolution answer would be an age difference (e.g. a $\sim$100~M$_\odot$ star could be born as an O star 
and evolve into WNLh/ha around an age of $\sim$1~Ma) but other possibilities (binarity, rotation, mergers, metallicity\ldots) exist. 

\subsection{HD~93\,129~A as a triple system}

$\,\!$\indent The analysis above is based on HD~93\,129~A having only two components. If HD~93\,129~Aa is a tight binary (as different pieces of evidence indicate), this 
periastron passage should be treated as a three-body problem and that opens up new possibilities. 

The first possibility is that one of the three stars is ejected at high speed and becomes a runaway star. That scenario for the 
production of a runaway star at the core of a compact cluster with massive stars was proposed almost half a century ago \citep{Poveetal67} but has never been observed in 
action (and probably never thought that it would take just decades instead of millenia to observe it for the first time). It is fitting that, if it indeed happens in this 
system, such a first occurrence takes place in Trumpler~14, which is the closest example of a compact massive young cluster with very massive stars.

A second possibility is a collision between two of the three stars if the periastron of the Aa,Ab orbit takes place at a distance comparable to the 
semi-major axis of the tight binary. Such an event between two massive stars has never been directly observed though eruptions in lower mass systems such as V838 Mon 
\citep{TyleSoke06} and V1309 Sco \citep{Tyleetal11} have been proposed (after the fact) as the result of stellar collisions (but possibly due to slow, tidally induced
inspiralling, not as the result of a fast, three-body interaction). Therefore, if the outcome of the periastron passage of the Aa,Ab orbit is the collision between either
pair of the three stars (Aa1, Aa2, and Ab) it would be the first time that [a] a stellar collision is followed through the process knowing the possible outcome
beforehand and [b] the event involves massive stars, one of them very massive. As it was the case with the first possibility, it would be fitting that such an event is 
first observed in the core of Trumpler 14, since the core of massive young stellar clusters has been proposed as the most likely location for them 
(see \citealt{Portetal99} for an optimistic version of the importance of stellar collisions and \citealt{MoecClar11} for a more pessimistic one).

A third possibility is that the periastron passage is relatively uneventful and Ab leaves it in a slightly perturbed orbit, returning decades or centuries 
later\footnote{A combination of these possibilities could also take place. E.g. the first Aa,Ab periastron passage could produce a serious orbital perturbation 
that would lead to one of the first two possibilities in a subsequent periastron passage.} (the
current uncertainty on $P$ is large, see Table~\ref{parameters}). Even if that is the case, the system is likely to be unstable in the long-term for three reasons:
[a] the current Aa,Ab eccentricity is very large, so even a small perturbation can cause a large effect; [b] as already mentioned, there are two nearby very massive 
stars, HD~93\,129~B and HD~93\,128 \citep{Sotaetal14}, and several other massive stars that may also perturb the already eccentric orbit; and [c] the Lidov-Kozai effect 
\citep{Koza62} can perturb the orbit of the inner pair and cause a collision between those two stars. Indeed, it has been proposed that $\eta$ Car was also a triple 
system and that the Lidov-Kozai effect caused two of the stars to collide and produce the mid-nineteenth eruption \citep{PortvadH16}.

Which one of those possibilities is more likely? We do not have an answer yet but the most important parameter that needs to be determined is the ratio between 
the Aa,Ab periastron distance and the Aa1,Aa2 semi-major axis. If that value is $\sim$1 or less (i.e. if the orbits cross or nearly so), then the first two possibilities 
can take place with relatively high probability. If, on the other hand, the value is $\gtrsim$3 we would be looking into the third scenario. The answer should come from
two sources: [a] further interferometric observations before the periastron passage to measure its distance more accurately and [b] a determination of the Aa1,Aa2 orbit
though a high-S/N multi-epoch spectroscopic campaign. We note the urgency of these observations: the chance to observe such a spectacular interaction between massive stars
may not be repeated in a long time. In that respect we have already obtained time for further PIONIER and spectroscopic observations in the incoming months.

What about the previous orbital evolution? The system may have been born in a similar configuration and Ab would have orbited around Aa thousands or tens of thousands of 
times with little variation from one orbit to another, along the lines of the third possibility above but going backwards in time. However, the three considerations listed
above (very high eccentricity, perturbations by other Trumpler 14 stars, and the Lidov-Kozai effect) make this unlikely: our impresion is that the system is more likely to 
have been born in a different configuration though a more definitive answer would require numerical simulations. This means that the number of previous orbits is currently
unknown. Going to the other extreme, if the $e \approx 0.991$ family of solutions turns out to be the correct one, Ab could have been recently captured by Aa and we could 
be witnessing not only one of the last but also one of the first periastron pasages of the system.

\section{Conclusions}

$\,\!$\indent The eccentric HD~93\,129~Aa,Ab system will pass through a close periastron in 2017/2018, providing a unique opportunity to study the interaction between two 
powerful stellar winds. The system is already an intense X-ray emitter but in a scenario with $d < 1$~mas, as suggested by our first family of orbital solutions,
the constraints we have placed on the relative orbit between the two components will produce a 
strong outburst at least an order of magnitude more intense than previous estimates. The evolution of the WCR properties should produce new insights into the particle
acceleration mechanisms.

We have also found three pieces of evidence that the Aa component is itself a binary system: [a] its spatially resolved spectrum shows features that are inconsistent with 
those of a single object, [b] the mass ratio derived from the absolute proper motions indicates that Aa is significantly more massive than Ab while the spectral 
classification points towards them being very similar objects, and [c] the spectral lines in the integrated spectrum show short-term velocity variations inconsistent with 
the long-term variations expected from the Aa,Ab orbit. Therefore, we expect the 2017/2018 periastron to be a three-body encounter with different possible outcomes. It 
could be uneventful in the short term if Ab passes at several times the Aa1,Aa2 distance but it could also lead to a runaway ejection or a stellar collision is the 
periastron distance is shorter. In any event, such an eccentric three-body system located at the center of a massive young cluster should be dynamically unstable in the 
long term and it is unlikely that it will remain unperturbed until the point where stellar evolution changes the characteristics of the stars significantly.

The comparison with $\eta$~Car is undoubtedly interesting, especially since $\eta$~Car may have been a triple system itself and its current status may have had more to do
with stellar dynamics than with stellar evolution. There are now two very massive multiple systems in the Carina Nebula with orbits so eccentric as to make their destinies
depart strongly from what we would expect for isolated stars or for nearly circular binary systems. How common are such systems? How influential are they for the 
evolution of very massive stars in general? What fraction of LBV eruptions are caused by three-body interactions? As usual in this field, better statistics are needed but 
the sample is likely to remain small in the short term.

\section*{Acknowledgements}

We thank the referee, Tony Moffat, for insightful comments that helped improve this paper.
We also thank N.~R.~Walborn for his inspiration to study HD~93\,129~Aa,Ab and E.~P.~Nelan for discussions on the astrometric motion
at an early stage of this project. The authors acknowledge the European Southern Observatory staff for their support.  
J.M.A. acknowledges support from the Spanish Government Ministerio de Econom{\'\i}a y Competitividad (MINECO) 
through grant AYA2013-40\,611-P and from HST GO programs 10\,205, 10\,602, and 10\,898. R.H.B. acknowledges support from FONDECYT Projects 
1\,140\,076 and 11\,121\,550. The HST data were obtained at the Space Telescope Science Institute, which is operated by the Association of 
Universities for Research in Astronomy, Inc., under NASA contract NAS 5-26555.




\bibliographystyle{mnras}
\bibliography{general} 



%
%


\bsp	
\label{lastpage}
\end{document}